\documentclass[aps,prx,twocolumn,showpacs,superscriptaddress,longbibliography]{revtex4-1}
\usepackage{graphicx}
\usepackage{array}
\usepackage{amsmath}
\usepackage[dvipsnames]{xcolor}
\usepackage[colorlinks=true, citecolor={blue!80!black}, urlcolor={blue!50!black}, linkcolor = {blue!80!black}]{hyperref}
\usepackage{amssymb}
\usepackage{bm}
\usepackage{color}
\usepackage{bookmark}
\usepackage{tabularx}
\usepackage{mathtools}
\usepackage{microtype}
\usepackage{hhline}
\usepackage{bbm}
\usepackage[normalem]{ulem}

\usepackage[percent]{overpic}

\newcommand{\Id}{\mathbbm{1}}
\newcommand{\vB}{v_{\rm B}}
\newcommand{\vBz}{v_{{\rm B},0}}
\newcommand{\lL}{\lambda_{\rm L}}
\newcommand{\ts}{t_*}

\begin{document}

\title{Scrambling and Lyapunov Exponent in Unitary Networks with Tunable Interactions}

\author{Anna Keselman}
\affiliation{Kavli Institute for Theoretical Physics, University of California, Santa Barbara, CA 93106-4030}
\affiliation{Microsoft Station Q, Santa Barbara, California 93106-6105, USA}

\author{Laimei Nie}
\affiliation{Kadanoff Center for Theoretical Physics, University of Chicago, Chicago, IL 60637, USA}
\affiliation{Department of Physics and Institute for Condensed Matter Theory, University of Illinois at Urbana-Champaign, Urbana, IL 61801, USA}

\author{Erez Berg}
\affiliation{Department of Condensed Matter Physics, Weizmann Institute of Science, Rehovot 76100, Israel}

\date{\today}
\begin{abstract}
Scrambling of information in a quantum many-body system, quantified by the out-of-time-ordered correlator (OTOC), is a key manifestation of quantum chaos. A regime of exponential growth in the OTOC, characterized by a Lyapunov exponent, has so far mostly been observed in systems with a high-dimensional local Hilbert space and in weakly-coupled systems.
Here, we propose a general criterion for the existence of a well-defined regime of exponential growth of the OTOC in spatially extended systems with local interactions. In such systems, we show that a parametrically long period of exponential growth requires the butterfly velocity to be much larger than the Lyapunov exponent times a microscopic length scale, such as the lattice spacing.
As an explicit example, we study a random unitary circuit with tunable interactions. In this model, we show that in the weakly interacting limit the above criterion is satisfied, and there is a prolonged window of exponential growth. Our results are based on numerical simulations of both Clifford and universal random circuits supported by an analytical treatment.
\end{abstract}

\maketitle

\emph{Introduction.-}
Many-body quantum chaos has recently attracted an increasing amount of attention thanks to its connections with quantum thermalization~\cite{Deutsch1991, Srednicki1994}, many-body localization~\cite{Basko2006MBL, PalHuse2010MBL}, and black hole physics~\cite{RobertsStanford2015diagnosing, ShenkerStanford2014butterfly, Maldacena2016bound, Cotler2017RMT}. Among the many operational diagnostics of quantum chaos~\cite{Hosur2016TOMI, Chen2018, Blake2020simulable, Nie2018TOMI, Jonay2018entanglement,Nahum2017entanglement,Jonay2018entanglement,Parker2019universal,Kos2018,Chan2018,Cotler2017,XiaoLudwig2018}, the scrambling of local quantum information, typically quantified by out-of-time-ordered correlators (OTOC)~\cite{Keyserlingk2018noconservation, Khemani2018conservation, Nahum2018spreading, Rakovszky2018conservation, Larkin1969,RobertsStanford2015diagnosing, ShenkerStanford2014butterfly, Maldacena2016bound,Aleiner2016combustion, Swingle2016measuring, Rozenbaum2017,gu2019relation},
aims to capture the growth of complexity of local operators under chaotic Heisenberg time evolution. In large-$N$ systems such as the Sachdev-Ye-Kitaev model and conformal field theories with large central charge and holographic duals, the OTOC has been shown to exhibit a regime of exponential growth characterized by the quantum Lyapunov exponent $\lL$~\cite{RobertsStanford2015diagnosing,Maldacena2016bound,Maldacena2016remarks, Chowdhury2017ONmodel, Fradkin2018Lifshitz}, serving as a probe for the chaotic dynamics in those systems. The scrambling time, which determines the time window for the exponential growth, is known to be parametrically long in the large-$N$ limit. 

OTOC in quantum many-body lattice systems with a finite-dimensional Hilbert space and local interactions has also been studied extensively, with rich structures observed in both early and late time regimes~\cite{DoraMoessner2017,Nahum2018spreading,Keyserlingk2018noconservation,Khemani2018conservation,XuSwingle2019,Prosen2017dOTOC,Luitz2017propagation}.
Nevertheless, in generic situations, no regime of exponential growth was found
~\cite{Nahum2018spreading,Keyserlingk2018noconservation,Luitz2017propagation}.
A special case, where it was shown that a regime of exponential growth is possible, is that of a weak coupling limit~\cite{Stanford2016,Patel2017,Aleiner2016combustion,Galitski2018unregularized,Schalm2019unregularized}. However, it remained unclear what controls the scrambling time in systems with a local structure.

In this work, we propose a general criterion for the existence of a well-defined time period of exponential growth of the OTOC, which is applicable for any system with spatial structure and local interactions. We argue that a parametrically long scrambling time in this setup can arise as a result of a competition between the exponential growth of the OTOC locally, and the rapid growth of the number of accessible degrees of freedom, which allow for a large enough phase space required for the exponential growth to persist. The latter growth rate is set by the butterfly velocity $\vB$~\cite{RobertsSwingle2016}, which is defined as the velocity of the propagation of the operator front upon the Heisenberg time evolution, and determines the size of the operator light cone. The existence of a parametrically long regime of exponential growth is thus possible in the limit of large $\vB/ (\lL a)$ ratio [see Eq.~\eqref{eq:crossover}], where $a$ is a microscopic length scale, to be discussed below. 

We demonstrate this principle using a random unitary circuit model, a minimally structured quantum system that has been employed to draw insights on the dynamical properties of deterministic quantum systems~\cite{Hayden2007mirror, Fawzi2015decoupling, Hosur2016TOMI, Nahum2018spreading, Rakovszky2018conservation, Keyserlingk2018noconservation, Khemani2018conservation, Nahum2017entanglement, Blake2020simulable}. Previous works on (1+1)D random unitary circuits showed that the front of the OTOC travels ballistically with a diffusive broadening, and no extended exponential regime with a fixed Lyapunov exponent was found~\cite{Nahum2018spreading, Keyserlingk2018noconservation}. 
Here, we introduce a random circuit model with a tunable parameter, that plays the role of interaction strength. We provide both analytical results and numerical verifications of the existence of an extended exponential regime in the OTOC in the limit of weak interaction. Our analysis further reveals the full structure of the OTOC during the entire evolution, including a crossover to a saturated regime at late times. In the limit of strong interactions, we recover the behavior observed in previous studies of Haar-random unitary circuits~\cite{Nahum2018spreading}.

Below, we start by defining the integrated OTOC, a quantity which we find to be more suitable to characterize operator growth in systems with spatial structure. 
We then introduce our random unitary circuit model, and show how the strength of the interactions can be tuned within it. Focusing on a special type of Clifford circuit, we demonstrate the existence of a regime of exponential growth, and characterize the crossover time to the saturated regime. Numerical results for this model are complemented by analytic rate equation which provide further insights on the temporal profile of the integrated OTOC. 
We then consider more generic Clifford and non-Clifford circuits, and show that our main results remain unchanged.

\begin{figure}
    \centering
    \begin{overpic}[width=0.49\columnwidth]{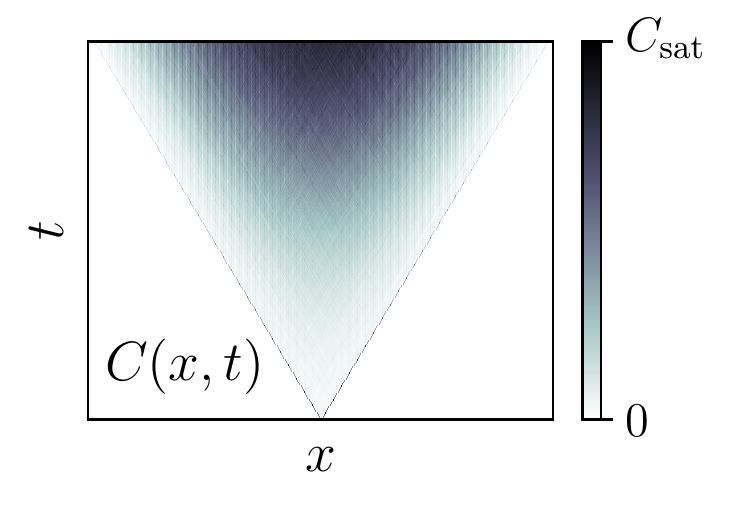} \put (0,68) {\footnotesize{(a)}} \end{overpic}
    \begin{overpic}[width=0.49\columnwidth]{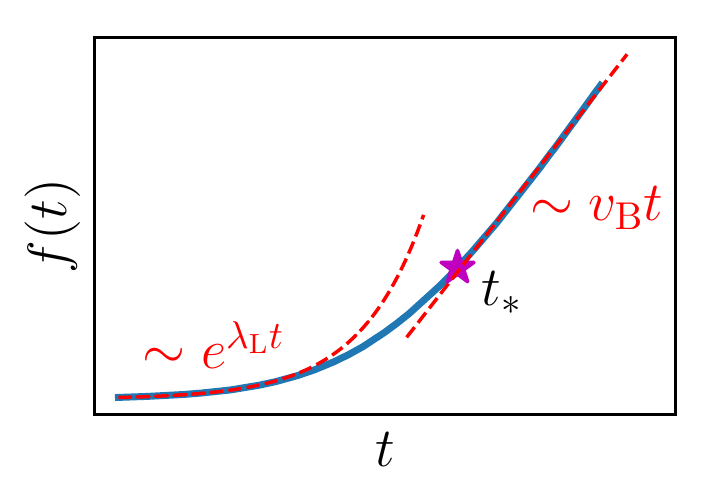} \put (0,68) {\footnotesize{(b)}} \end{overpic}
    \caption{Time evolution of the (a) local and (b) integrated OTOC in the special Clifford circuit discussed in the text, demonstrating the generic behavior expected in a one-dimensional system with local interactions. Early time growth is characterized by a Lyapunov exponent $\lL$, while at late times the growth of the iOTOC is linear, with a rate determined by the butterfly velocity $\vB$ and the saturation value of local OTOC, $C_{\rm sat}$. The crossover time between the two regimes (the scrambling time) is indicated by $\ts$.}
    \label{fig:Schematic}
\end{figure}

\emph{Scrambling time in systems with local interactions.-}
We consider a finite-dimensional system defined on a lattice, where each lattice site contains a degree of freedom with a finite-dimensional Hilbert space.
The OTOC of two local operators, $W_i,V_j$, acting on sites $i,j$ respectively, is given by
\begin{equation}
C_{i,j}(t) = - \left\langle  \left[W_i(t),V_j\right]^2\right\rangle.
\label{eq:localOTOC}
\end{equation}
Since in this work we focus on a random unitary circuit evolution, for which there is no well-defined notion of temperature, we take the expectation value above to be over an ensemble of random states, corresponding to the infinite temperature limit.
While our arguments hold for systems of an arbitrary finite spatial dimension, for simplicity, we will focus on the one-dimensional case below and address higher dimensions later on.
In a generic scenario, upon time evolution, the support of the operator $W_i(t)$ grows ballistically, forming a light cone with the front propagating at the butterfly velocity $\vB$. The OTOC above becomes finite once the site $j$ enters this light cone. 
Following an early exponential growth regime, the value of $C_{i,j}(t)$ must saturate at late times, as its value is bounded due to the finite dimension of the local Hilbert space.
This behavior is shown in Fig.~\ref{fig:Schematic}(a), for a random circuit model to be described below. 

While the structure and dynamics of the local OTOC~\eqref{eq:localOTOC} are interesting on their own right, here we would like to focus on the global properties of the scrambling dynamics.
To this end, we introduce the integrated OTOC (iOTOC),
\begin{equation}
f(t)=\sum_j C_{i,j}(t) = - \sum_j\left\langle \left[W_i(t),V_j\right]^2\right\rangle,
\label{eq:OTOC}
\end{equation}
where a summation is performed over all the lattice sites of the system~\footnote{A closely related quantity has been considered in the past in Ref.~\cite{Prosen2017dOTOC}. There, the focus was on the late-time behavior of the OTOC.}.
The iOTOC measures the expectation value of the ``size" of the operator~\cite{Roberts2018growth,lucas2020non}, washing out any transient behaviors and details of spatial structure, and thus simplifying the identification and characterization of the scrambling time in systems with local structure. 
Similarly to the local OTOC, at early times, we expect the iOTOC to exhibit an exponential growth characterized by a Lyapunov exponent $f\left(t\right)\sim e^{\lL t}$. At late times, when the OTOC in the bulk of the system reaches its saturation value $C_{\rm sat}$, the iOTOC crosses over to a linear growth regime (due to the linear growth of the size of the light cone), $f(t)\sim C_{\rm sat} \vB t $.
Assuming a single crossover time $\ts$ (the scrambling time) between these two regimes, it can be obtained from
\begin{equation}
e^{\lL \ts} \sim c_{\rm sat} \vB \ts \Rightarrow \ts \sim\frac{1}{\lL}\log\frac{c_{\rm sat}\vB}{\lL},
\label{eq:crossover}
\end{equation}
where we introduced the OTOC density at saturation $c_{\rm sat}=C_{\rm sat}/a$, with $a$ being the lattice spacing.
We thus find, that a parametrically long scrambling time is expected for a diverging ratio $\vB/\lL$, allowing for a prolonged time window where an exponential growth regime could be observed in systems with local structure.
The iOTOC, corresponding to the local OTOC shown in Fig.~\ref{fig:Schematic}(a), is plotted in Fig.~\ref{fig:Schematic}(b), where this behavior can be observed.

\begin{figure}
    \centering
    \begin{overpic}[width=0.49\columnwidth]{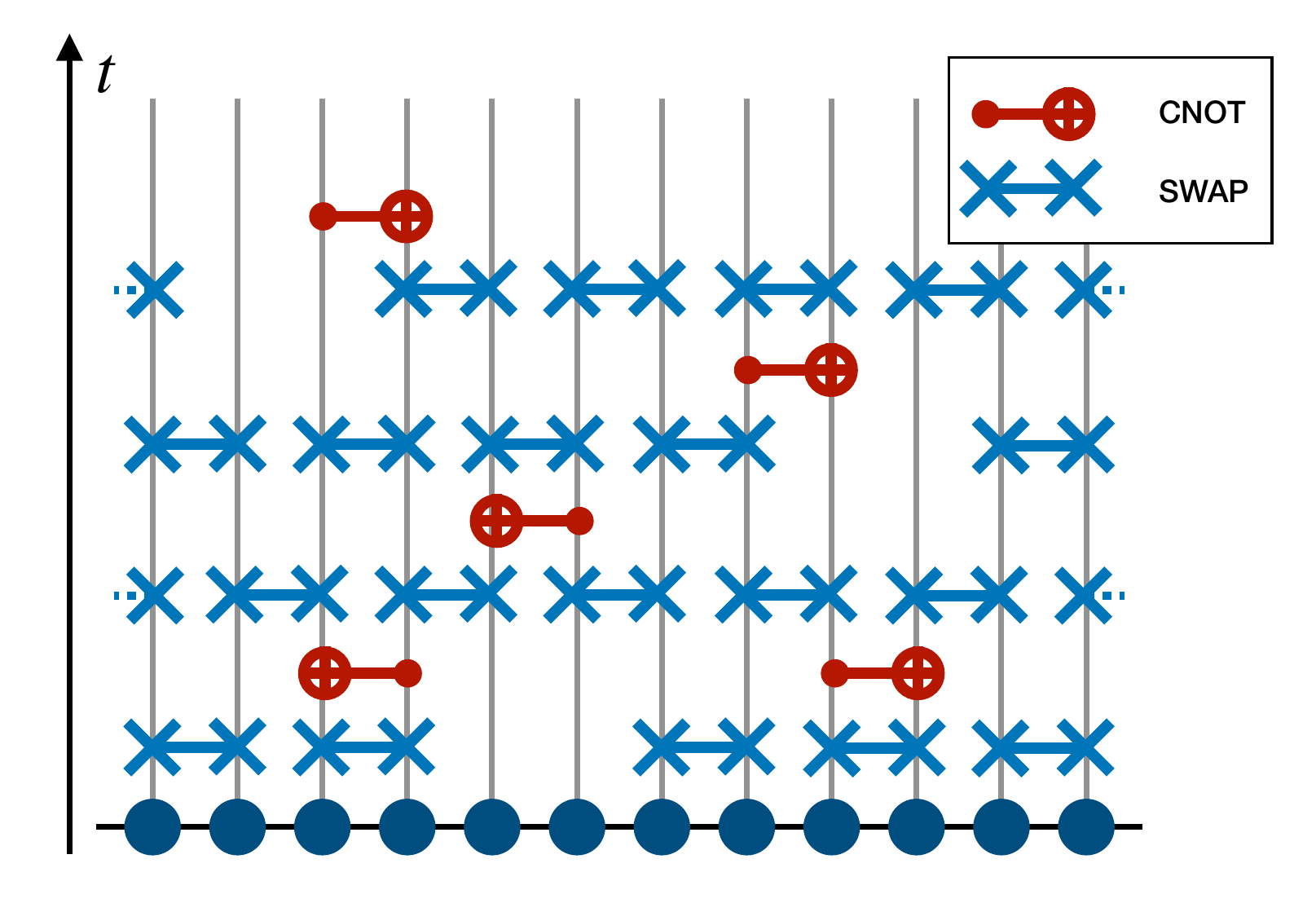} \put (0,68) {\footnotesize{(a)}} \end{overpic}
    \begin{overpic}[width=0.49\columnwidth]{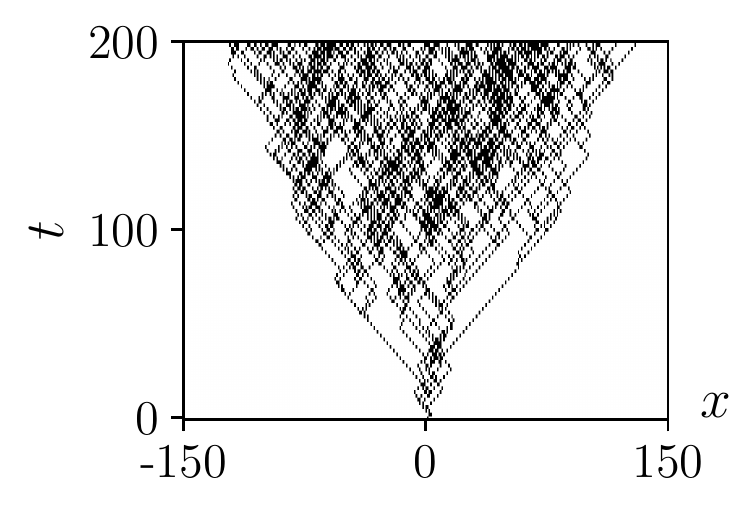} \put (0,68) {\footnotesize{(b)}} \end{overpic}
    \caption{(a) Schematic representation of the Clifford unitary circuit. At odd (even) time steps a set of SWAP gates is applied on odd (even) bonds with probability $p$ on each bond, followed by a set of CNOT gates which are applied on a fraction $r$ of the bonds. (b) Density of $\sigma^z$ operators in the operator string $\sigma^z_0(t)$, for a single realization of the circuit with $p=0.9,r=0.05$.
}
    \label{fig:circuit}
\end{figure}

\emph{Random Clifford circuit model.-}
To demonstrate the arguments above, we study the scrambling dynamics in a random unitary circuit.
We start with the simplest circuit model in which the general behavior discussed above can be observed, and which is amenable to both analytical and large scale numerical analysis. A study of a more generic version of the circuit, which concurs with the results obtained here, is presented later on.
The structure of the circuit we consider here is shown schematically in Fig.~\ref{fig:circuit}(a).
At every odd (even) time step a set of SWAP gates is applied on the odd (even) bonds, with probability $0 \leq p\leq1$ on each bond. 
A SWAP gate interchanges the state of the two qubits it acts on, and can be written explicitly as $\sigma_1^+\sigma_2^- + \sigma_1^-\sigma_2^+ + (1+\sigma_1^z\sigma_2^z)/2$, where $\sigma^z$ and $\sigma^\pm$ are Pauli operators.
Then, a set of CNOT gates is applied on a fraction $0 \leq r \leq 1/2$ of all the bonds. The bonds are chosen such that only configurations where no two bonds share a site are allowed and each such configuration is equally probable.
The role of each qubit (control or target) is chosen randomly and independently for each pair of sites.

First, note that the circuit consists of Clifford gates only and therefore can be simulated classically~\cite{Gottesman1998}, allowing us to explore large system sizes and long times. 
Upon Clifford evolution, an operator $\sigma^{\alpha}_i(t=0)=\sigma^{\alpha}_i$, with $\alpha=x,y,z$,
remains a single operator string of Pauli operators.
In particular, for the circuit structure described above, when the operator $\sigma^{z}_i(t)$ is considered, the corresponding operator string at times $t>0$ consists only of $\sigma^z$ and identity operators, as we will see below.
Fig.~\ref{fig:circuit}(b) shows the density of $\sigma^z$ operators in this case as a function of time, for a single realization of the circuit.

To better understand the evolution of operators in this circuit consider first the limit of $r=0$ (i.e. in the absence of CNOT gates), and $p=1$ (i.e. when the SWAP gates are applied on all odd / even bonds at each time step). 
In this case, a single-site operator located on an odd (even) site of the 1D chain, will propagate ballistically to the right (left) with velocity $\vBz=1$.
Decreasing $p$ can be thought of as introducing disorder, as a missing SWAP gate results in a back-scattering of an operator, flipping its velocity. This gives rise to a diffusive propagation with diffusion constant $D\sim \vBz^2 \tau_p$, where $\tau_p\sim (1-p)^{-1}$ is the characteristic time between consecutive back-scatterings.

Next, consider a finite $r>0$, and for concreteness focus on the evolution of a $\sigma^z$ operator. The action of a CNOT gate on operators (when the first qubit is the control qubit and the second one is the target) is given by
\begin{equation}
\sigma^z \otimes \Id \to \sigma^z \otimes \Id,\ 
\Id \otimes \sigma^z \to \sigma^z \otimes \sigma^z,\ 
\sigma^z \otimes \sigma^z \to \Id \otimes \sigma^z.
\label{eq:CNOT}
\end{equation}
This process can be thought of as a scattering event for the operators due to interactions, which increases the support of a local operator. 
Denoting the scattering time due to the CNOT gates as $\tau_r\sim r^{-1}$, we note that in the diffusive case ($p<1$), a finite $r$ gives rise to a finite butterfly velocity $\vB\sim\sqrt{D/\tau_r}\sim\sqrt{r/(1-p)}$~\cite{Patel2017}.

We are interested in the evolution of the OTOC in this circuit.
Writing the operator string corresponding to $\sigma^{\alpha}_i(t)$ explicitly as $\otimes_k\sigma^{\alpha_k(t)}_k$ (where $k$ runs over all the sites of the 1D system, and we use $\sigma^0$ to denote the identity), the commutator $\left[\sigma^{\alpha}_i(t),\sigma^{\beta}_j\right]$ is given by 
$\left( \otimes_{k\neq j}\sigma^{\alpha_k(t)}_k \right) \otimes\left[\sigma^{\alpha_j(t)}_j,\sigma^{\beta}_j\right]$. Since $\left(\sigma^{\alpha_k(t)}_k\right)^2=\Id$ for any $\alpha_k(t)$, the commutator squared is simply $\left[\sigma^{\alpha_j(t)}_j,\sigma^{\beta}_j\right]^2=4(\delta_{\beta,\alpha_j(t)}-1)\Id$ for $\alpha_j(t)\neq 0$ (i.e. $\sigma_j^{\alpha_j(t)}\neq\Id$). In particular, the expectation value is state independent.
Performing the summation over $j$ in Eq.~\eqref{eq:OTOC} with $W_i=\sigma^\alpha_i/\sqrt{2}$ and $V_j=\sigma^\beta_j/\sqrt{2}$, we find $f(t)=\sum_j (1-\delta_{\beta,\alpha_j(t)})(1-\delta_{\alpha_j(t),0})$, i.e. the number of (non-identity) Pauli operators in $\sigma^{\alpha}_i(t)$ which are different from $\sigma^\beta$.
For concreteness, below, we will consider the OTOC between $\sigma^z_i(t)$ and $\sigma_j^x$ , so that $f(t)$ amounts to the number of $\sigma^z$ operators in the string $\sigma^z_i(t)$ at time $t$. Note that the commutator $[\sigma^z_i(t), \sigma_j^z]$ vanishes identically for any realization of the circuit in this model. In fact, any state which is a product state in the $z$ basis remains un-entangled upon evolution with the circuit above. However, this property of the circuit is not important for the discussion below, and we will show later on that our results hold for the more generic case as well.

\begin{figure}
    \centering
    \begin{overpic}[width=0.54\columnwidth]{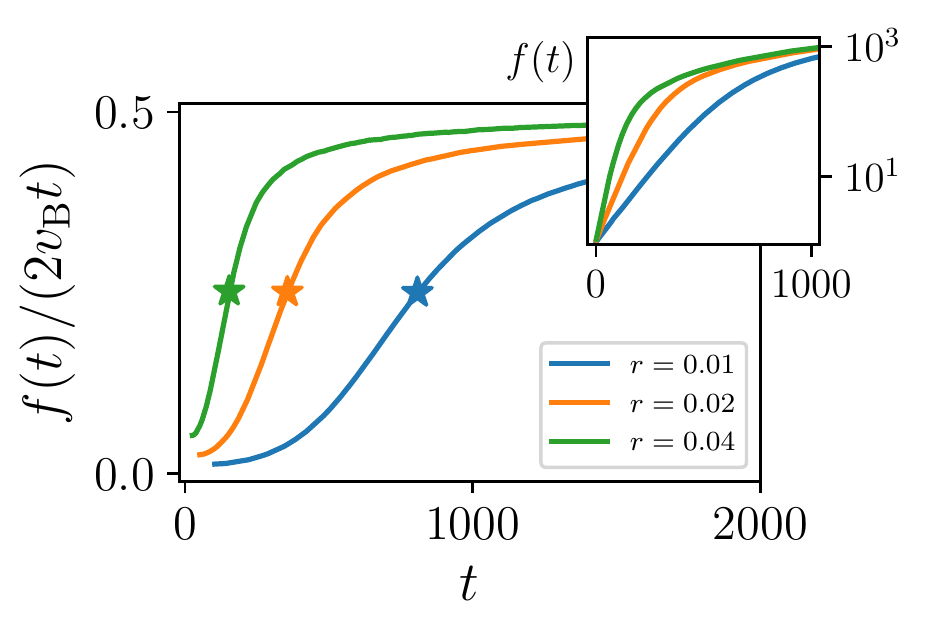} \put (0,60) {\footnotesize{(a)}} \end{overpic}
    \begin{overpic}[width=0.44\columnwidth]{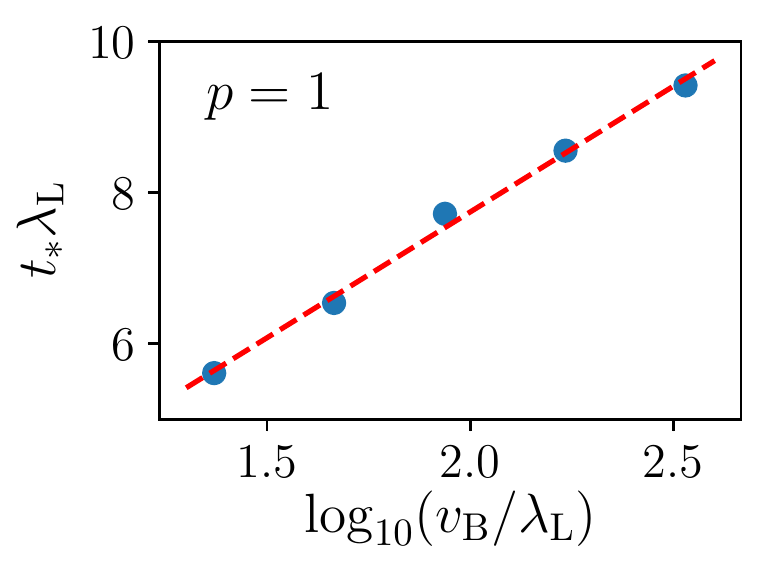} \put (0,73) {\footnotesize{(b)}} \end{overpic} \\
    \begin{overpic}[width=0.54\columnwidth]{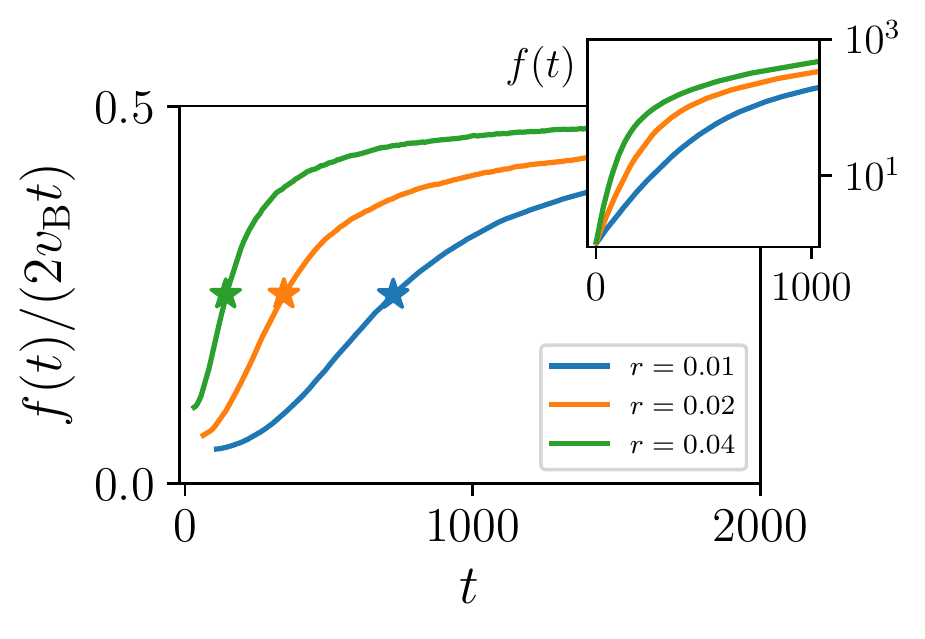} \put (0,60) {\footnotesize{(c)}} \end{overpic}
    \begin{overpic}[width=0.43\columnwidth]{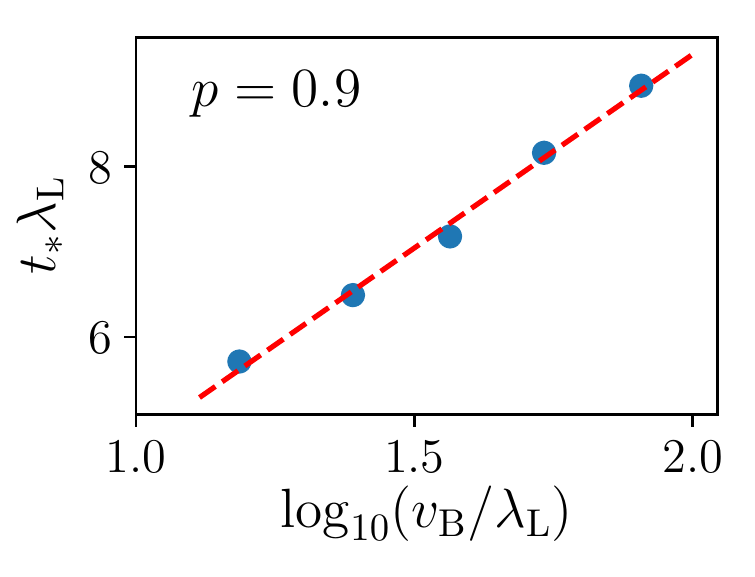} \put (0,74) {\footnotesize{(d)}} \end{overpic}
    \caption{Numerical results for the OTOC in the Clifford circuit, for the ballistic case ($p=1$) in (a,b), and for the diffusive case ($p=0.9$) in (c,d). (a,c) The OTOC density, $f(t)/(2\vB t)$, for different values of $r$ as function of time. The crossover time, $\ts$, when the OTOC density reaches half of the saturation value, is indicated by a star. Insets: $f(t)$ on a log scale. (b,d) Scaling of $\ts$ with $\lL$ (extracted as the slope of the integrated OTOC at early times) and $\vB$ (extracted from the spatial profile of the operator density at late times). 
    }
    \label{fig:OtocsClifford}
\end{figure}

\emph{Numerical results for integrated OTOCs and crossover time.-}
We now study the behavior of the iOTOC in this model as function of the circuit parameters.
To this end, we perform numerical simulations of the operator dynamics, calculating the local and integrated OTOC.
As was already mentioned, the density of the CNOT gates, $r$, is the parameter that sets the growth rate of the support of a local operator in the circuit and leads to scrambling. Therefore, we expect the Lyapunov exponent $\lL$ to be directly determined by $r$.
In the limit $r\ll 1$ (and hence large $\vB/\lL)$, we expect an extended time regime in which the growth of the iOTOC is exponential with a well-defined Lyapunov exponent. We find that this is indeed the case both for the ballistic and the diffusive parameter regime, as can be seen in the insets of Figs.~\ref{fig:OtocsClifford}(a,c).
To further analyze the crossover time, and its scaling with $\lL$ and $\vB$, we look at the average OTOC density, namely the iOTOC, $f(t)$, divided by the size of the light cone, $2\vB t$ (see Figs.~\ref{fig:OtocsClifford}(a,c)). At late times we expect this quantity to approach the saturation value of the OTOC in the bulk, which we find to be $1/2$ and independent of $r$ in the regime $r\ll 1$. (This value is in agreement with the expectation from the analytic rate equation analysis presented later on.)
We define the crossover time $\ts$ as the time at which the averaged OTOC density reaches half of its saturation value. In Figs.~\ref{fig:OtocsClifford}(b,d) we show that the crossover time extracted as above, indeed obeys the scaling expected from Eq.~\eqref{eq:crossover} both for the ballistic circuit with $p=1$ and the diffusive one with $p=0.9$. The results were obtained by averaging over $4\cdot10^3$ ($10^4$) realizations for the ballistic (diffusive) case.

\emph{Master equation for integrated OTOC.-}
To gain further insights on the scrambling process in the model described above, we now derive analytic rate equations for the iOTOC.
We consider the limit of small but finite $r$, and $1-p\ll 1$, such that the scattering events due to the CNOT gates are dilute and can be assumed to be uncorrelated~\cite{SM}.
This assumption is analogous to the molecular chaos hypothesis.

At time step $t$, the number of CNOT gates applied within the light cone of an operator $\sigma_i^z(t)$ is $N_{\rm CNOT}=2r\vB t$. Consider what happens to the total number of $\sigma^z$ operators in the operator string upon application of a CNOT gate. From~\eqref{eq:CNOT}, we see that this number increases by one if the target (but not the control) site hosts a $\sigma^z$ operator, while if both sites host a $\sigma^z$ operator, the number decreases by one.
Denote the fraction of non-identity operators in the operator string of $\sigma_i^z(t)$, within its light cone, by $q\equiv f(t)/(2\vB t)$. Assuming the probabilities of different sites to host a $\sigma^z$ operator are independent, the probabilities for the processes which increase or decrease the number of non-identity operators in the string are given by $q(1-q)$ and $q^2$, respectively.
Thus, the change in the number of $\sigma^z$ operators in the operator string in a single time step is given by $N_{\rm CNOT}\left(q(1-q)-q^2\right)$.
Recalling that the iOTOC is given simply by the number of non-identity operators in the operator string, as discussed above, we find that the rate equation for the iOTOC is (treating the time as continuous)
\begin{equation}
\frac{df}{dt}= r f(t)\left(1-\frac{f(t)}{\vB t}\right).
\label{eq:master_eq}
\end{equation}
This equation admits a solution of the form 
\begin{equation}
f(t)= \frac{g_0 e^{rt}}{1+g_0\frac{r}{\vB}\left[ {\rm Ei}(rt) - {\rm Ei}(1) \right] },
\label{eq:master_sol}
\end{equation}
where ${\rm Ei}(rt)$ is the exponential integral, and $g_0$ is a constant set by the initial conditions.
At early times, we see that indeed $f(t)\sim e^{\lL t}$, with a Lyapunov exponent set by the CNOT gates density, $\lL=r$. At late times, using the asymptotic expansion for the exponential integral, we find $f(t)\simeq \vB t / (1+(rt)^{-1})$, i.e. the slope asymptotically approaches the butterfly velocity. The average OTOC density, $f(t)/(2\vB t)$ thus tends to $1/2$ as observed numerically (see Figs.~\ref{fig:OtocsClifford}(a,c)).
The crossover time, $\ts$, at which the OTOC density reaches a finite fraction of the saturation value, is given (to leading order) by $e^{r \ts}\sim \vB \ts $ in agreement with Eq.~\eqref{eq:crossover} and as observed numerically.
Although the focus of our discussion here was on the iOTOC, in which the spatial structure is washed out, in the Supplementary Material (SM)~\cite{SM} we discuss hydrodynamic equations capturing the spatial structure and discuss their validity. We observe a crossover from a propagation in which the front maintains its shape to a regime where the front broadens diffusively.

\begin{figure}    
    \begin{overpic}[width=0.5\columnwidth]{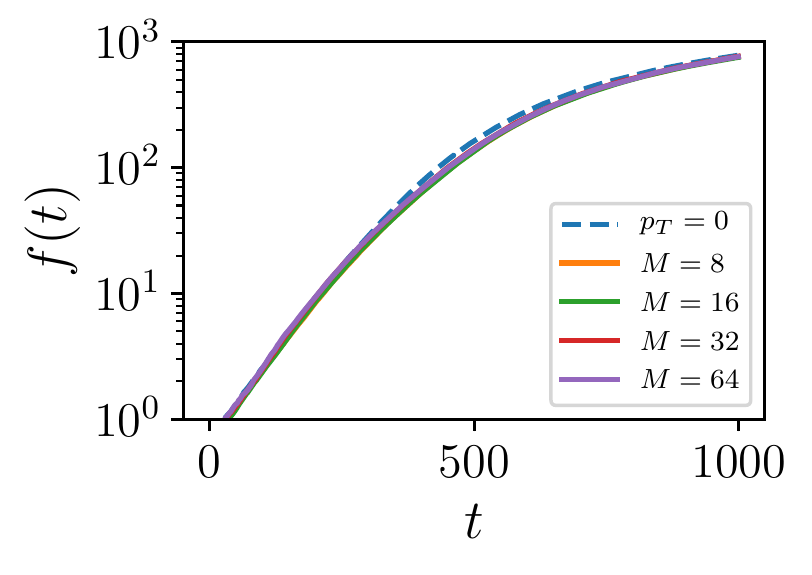} \put (0,67) {\footnotesize{(a)}} \end{overpic}
    \begin{overpic}[width=0.48\columnwidth]{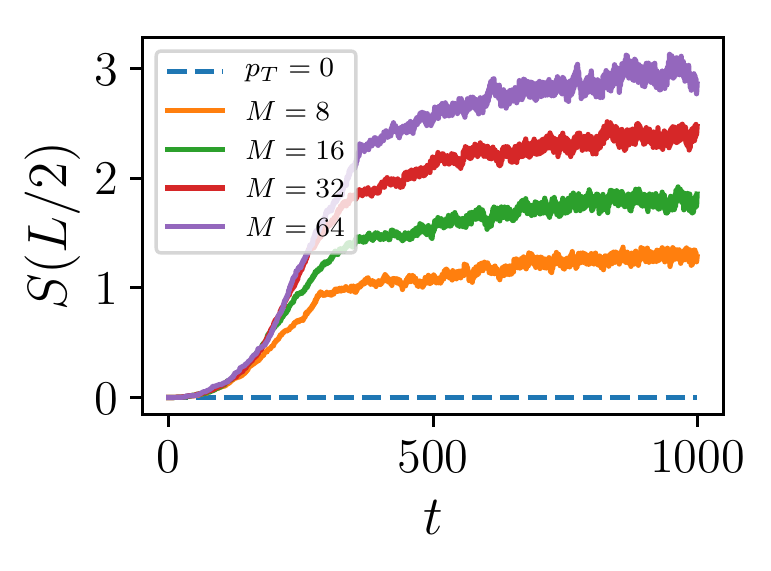} \put (0,70) {\footnotesize{(b)}} \end{overpic}
    \caption{(a) Integrated OTOC obtained for the generalized circuit models, shown on a log-scale, and (b) operator entanglement entropy across the middle bond in the system, as function of time, for different bond dimensions $M$. Averaging over $300$ realizations of the circuit is performed in each case. The dashed line corresponds to the generalized Clifford evolution, in the absence of T gates (in this case operator entanglement remains zero at all times). 
    }
    \label{fig:OtocsWithT}
\end{figure}

\emph{Generalizations of the random circuit model.-}
As noted previously, the circuit model considered above is a  special type of a Clifford circuit, in which both operator entanglement and state entanglement do not grow upon time evolution. We now demonstrate that our results do not rely on either of this properties.
For simplicity, we restrict the analysis below to the ballistic case, i.e. $p=1$.

First, consider a generalization of the circuit, in which the standard CNOT gate is replaced by a CNOT operation where the basis for both the control and the target qubits is chosen to be the x,y or z basis randomly and independently for each of the two qubits.
Namely
\begin{equation}
U^{\alpha,\beta}_{CNOT} = \frac{\Id_1+\sigma^\alpha_1}{2}\otimes\Id_2 + \frac{\Id_1-\sigma^\alpha_1}{2}\otimes e^{i\frac{\pi}{4}\sigma^\beta_2},
\label{eq:gen_cnot}
\end{equation}
where $\alpha,\beta\in {x,y,z}$.
While this remains a Clifford circuit, the entanglement of a state generically grows with time in this case. We consider the averaged OTOC $\propto\sum_{\mu,\nu=x,y,z} \left\langle[\sigma^\mu_i(t),\sigma_j^\nu]^2\right\rangle$.
The averaged iOTOC obtained for this model, for CNOT gate density $r=0.01$, is plotted in Fig.~\ref{fig:OtocsWithT}(a) (dashed blue line). It can be seen that a prolonged regime of exponential growth is present, similarly to the simplified model. Additional results for this model, and in particular a verification of the scaling in Eq.~\eqref{eq:crossover}, are given in the SM~\cite{SM}.

We next consider a further generalization to a non-Clifford circuit. To this end, at each time step of the evolution, following the application of CNOT gates, a T gate (i.e. a $\pi/4$ phase gate, around a randomly chosen axis) is applied with probability $p_T$ at each site. 
To calculate the OTOCs in presence of T gates we perform the time evolution of operators using a matrix product state (MPS)~\cite{Schollwoeck2011} representation of the operator string, employing the ITensor library~\cite{ITensor}.
Due to the exponential growth of operator entanglement exact simulations are limited to short times. To go to longer times we perform truncation of the MPS bond dimension. 
In Fig.~\ref{fig:OtocsWithT}(a) we plot the iOTOC for $r=0.01$ and T gate density $p_T=0.01$, for different maximal bond dimensions. The respective operator entanglement that builds up in the system is shown in Fig.~\ref{fig:OtocsWithT}(b). We see that although the operator entanglement in the system is now non-zero, the iOTOCs are essentially unmodified. Note that evolution up to times $t \sim 300$ is carried out without any truncation, and is thus exact.

\emph{Discussion.-}
In this work we proposed a condition for the existence of a Lyapunov exponent in many-body systems with local interactions and a finite dimensional on-site Hilbert space. Having a parametrically long scrambling time (where the OTOC is exponentially growing, and hence $\lL$ is well-defined) requires the ratio $v_B/\lL$ to be large. This condition is naturally fulfilled in weakly coupled systems. Whether the condition is satisfied in other situations, e.g., in generic strongly-coupled systems in the low-temperature limit, remains to be seen.

Our condition is demonstrated in an explicit one-dimensional random unitary circuit model, where we have verified the relation between the scrambling time and $v_B/\lL$. However, we expect the results to carry over to higher dimensions. Since the number of sites in the light cone grows as $(\vB t)^d$ in the d-dimensional case, the late-time iOTOC scales as $f(t)\sim t^d$. Therefore, the parametrically large scrambling time is enhanced by a factor of $d$ relative to the one-dimensional case. 

Finally, we note that other probes for scrambling have been proposed, in particular the growth of state and operator entanglement~\cite{Hosur2016TOMI, Nahum2018spreading, Chen2018, Blake2020simulable, Keyserlingk2018noconservation, Nie2018TOMI, Jonay2018entanglement,Nahum2017entanglement,Jonay2018entanglement}. In our Clifford circuit, we find an exponential growth of the OTOC despite the fact that the operator entanglement (as well as the state entanglement in the special circuit described above) do not grow,  indicating that the existence of a Lyapunov exponent captures a different aspect of scrambling.

\begin{acknowledgments}
\emph{Acknowledgements-} 
We thank Ehud Altman, Bela Bauer, Eduardo Fradkin, Dima Pikulin, Steve Shenker, and Brian Swingle for stimulating discussions.
This research is funded in part by the Gordon and Betty Moore Foundation through Grant GBMF8690 to UCSB to support the work of A.K. in KITP. Use was made of the computational facilities administered by the Center for Scientific Computing at the CNSI and MRL (an NSF MRSEC; DMR-1720256) and purchased through NSF CNS-1725797. L.N. is supported by the Kadanoff Fellowship at the University of Chicago and NSF grant DMR-1725401 at the University of Illinois.
EB was supported by the European Research
Council (ERC) under grant HQMAT (grant no. 817799), by the Israel Science Foundation Quantum Science and
Technology grant no. 2074/19, and by
CRC 183 of the Deutsche Forschungsgemeinschaft.
We thank the hospitality of the Kavli Institute for Theoretical Physics, supported by the NSF under Grant No. NSF PHY-1748958, and the Aspen Center for Physics, supported by NSF grant PHY-1607611, where parts of this work were performed.
\end{acknowledgments}

\bibliographystyle{unsrt}
\bibliography{References_ScramblingRandomUnitaries.bib}

\clearpage

\setcounter{equation}{0}
\setcounter{figure}{0}
\renewcommand{\theequation}{S\arabic{equation}}
\renewcommand{\thefigure}{S\arabic{figure}}

\renewcommand{\thesection}{S\arabic{section}}
\renewcommand{\thesubsection}{\thesection.\arabic{subsection}}
\renewcommand{\thesubsubsection}{\thesubsection.\arabic{subsubsection}}

\onecolumngrid

\begin{center}
{\Large\bfseries Supplementary Material}
\end{center}

\twocolumngrid

\section{Generalized Clifford circuit}\label{app:clifford}
\begin{figure}[h]
    \centering
    \begin{overpic}[width=0.49\columnwidth]{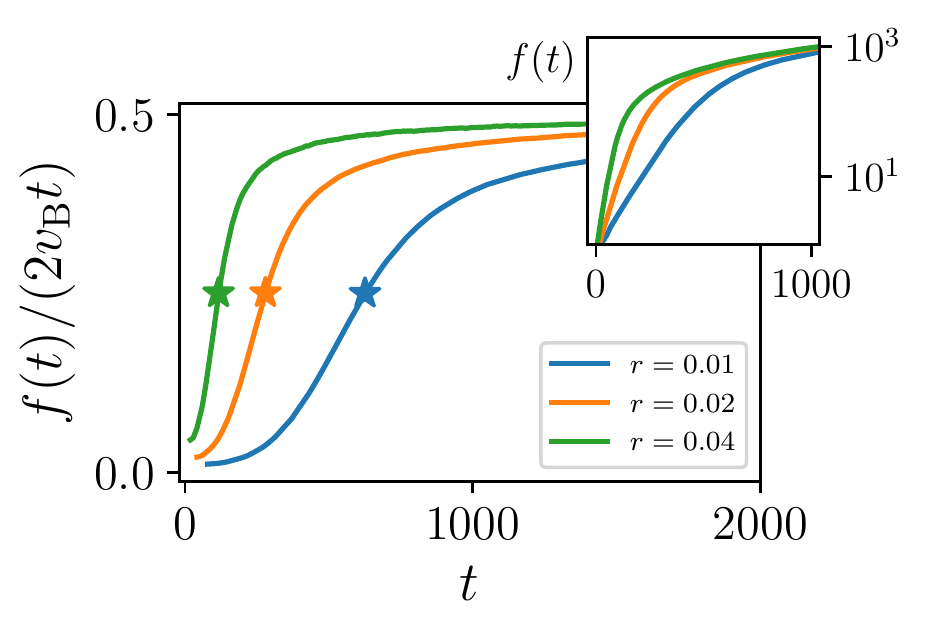} \put (0,60) {\footnotesize{(a)}} \end{overpic}
    \begin{overpic}[width=0.49\columnwidth]{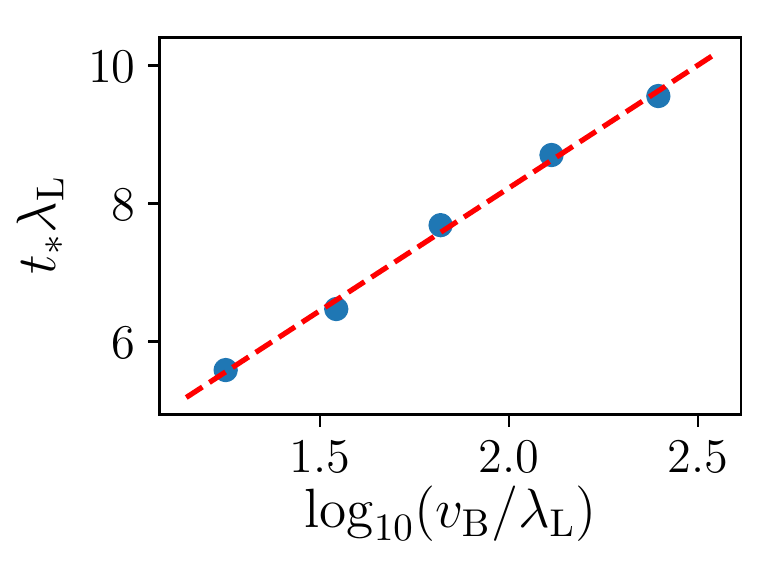} \put (0,60) {\footnotesize{(b)}} \end{overpic}
    \caption{(a) The OTOC density, i.e. $f(t)/(2\vB t)$, for different values of $r$ in the generalized Clifford circuit with $p=1$. Crossover time at which the OTOC density reaches half of its saturation value is marked by a star. The inset shows the iOTOC on a log scale. (b) Scaling of the crossover time expected from Eq.~\eqref{eq:crossover} in the main text. }
    \label{fig:gen_clifford}
\end{figure}

In Fig.~\ref{fig:gen_clifford} we present additional results for the generalized Clifford circuit, in which CNOT gates are replaced by two-qubit operators defined in Eq.~\eqref{eq:gen_cnot} in the main text. Here, an average over $400$ circuit realizations was performed, and an averaged OTOC over the different pauli operators, i.e. $\propto\sum_{\mu,\nu=x,y,z} [\sigma^\mu_i(t),\sigma_j^\nu]^2$ was calculated.
It can be seen that the behavior of the iOTOC in this case is similar to the one observed for the special Clifford circuit in Fig.~\ref{fig:OtocsClifford} in the main text.
In particular, a prolonged regime of exponential growth can be clearly seen for small values of $r$, and the scaling of the crossover time, expected from Eq.~\eqref{eq:crossover} in the main text, holds.

\section{Spatial structure of the OTOC and hydrodynamic equations for $C(x,t)$}\label{app:spatial}
In the main text we focused on the integrated OTOC, arguing that this quantity allows for an easier identification and characterization of the scrambling time in systems with local structure. For completeness, here we discuss the spatial structure of the OTOC in the circuit under investigation.

\subsection{Local OTOCs}

\begin{figure}[h]
    \centering
    \begin{overpic}[width=0.48\columnwidth]{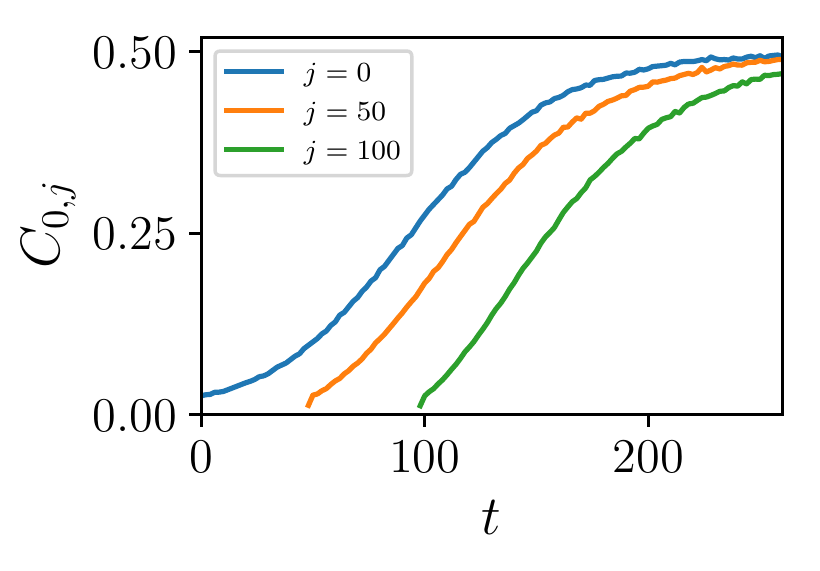} \put (0,60) {\footnotesize{(a)}} \end{overpic}
    \begin{overpic}[width=0.49\columnwidth]{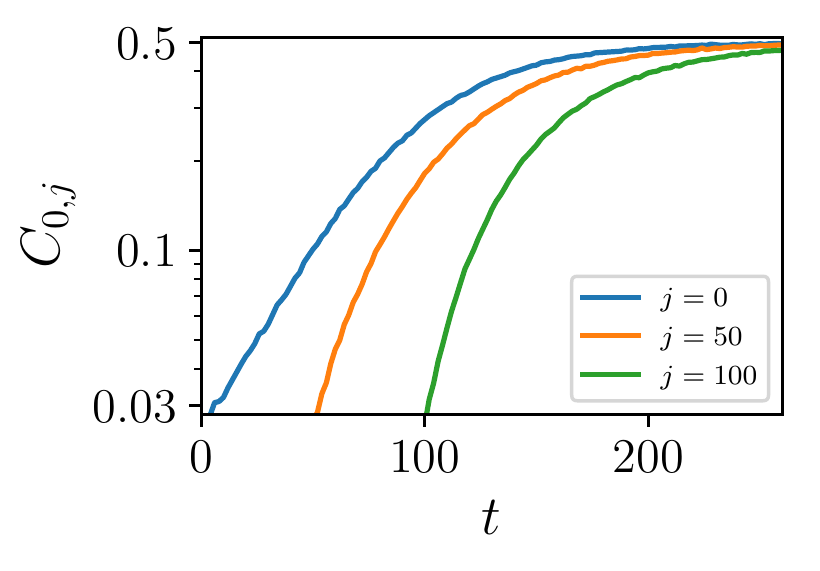} \put (0,60) {\footnotesize{(b)}} \end{overpic}
    \caption{Local OTOC $C_{0,j}$ probed at different values of $j$ on a linear scale in (a) and a log scale in (b), obtained for the special Clifford circuit with $p=1$ and $r=0.05$. }
    \label{fig:local_otoc}
\end{figure}

In Fig.~\ref{fig:local_otoc}, we show the local OTOC $C_{i,j}(t)=\left\langle[\sigma^z_i(t),\sigma^x_j]^2\right\rangle$ as function of time, at different positions along the 1D chain, obtained for the special Clifford circuit with $p=1$ and $r=0.05$. For $j=i$, an exponential growth of the OTOC at early times can be clearly seen. For $j\neq i$ the OTOC vanishes identically before the arrival of the light cone at time $t=\vB |j-i|$. While for $|j-i|\ll \ts/\vB$ we expect a regime of exponential growth to be present, for $|j-i| \gtrsim \ts/\vB$ this is no longer the case as can be seen in the Figure.

\subsection{Scaling of $\vB$ with $\lL$ in the diffusive circuit}

\begin{figure}[h]
    \centering
    \includegraphics[width=0.5\columnwidth]{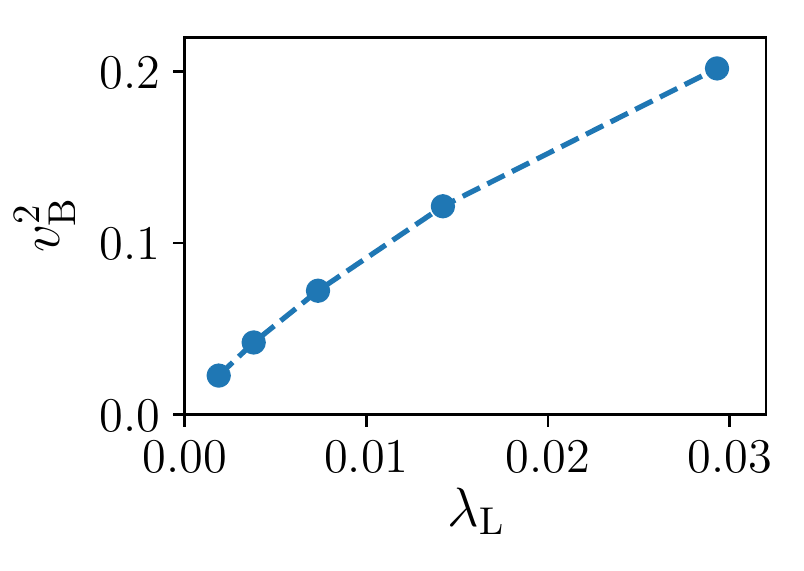}
    \caption{Scaling of the butterfly velocity $\vB$ with the scattering rate $\lL$ for a diffusive Clifford circuit with $p=0.9$.}
    \label{fig:vB_lL}
\end{figure}

In the main text, we argued, that to allow for a parametrically large regime of exponential growth of the OTOC, a diverging ratio of $\vB/\lL$ is required.
When the dynamics in the absence of interactions is diffusive (which in the case of the random circuit model under study occurs when $p<1$), a finite scattering rate (i.e. a finite $\lL$) is required to generate a a well-defined butterfly velocity, $\vB$.
It is thus important to verify the scaling of $\vB$ with $\lL$ still allows for a diverging ratio $\vB/\lL$ in the $\lL\to0$ limit. In Fig.~\ref{fig:vB_lL} we show that in the small $\lL$ regime, the scaling is $\vB\propto\sqrt{\lL}$ as expected~\cite{Patel2017}.

\subsection{Hydrodynamic equations for $C(x,t)$}

\subsubsection{Regime of validity of the hydrodynamic equations}
In deriving the hydrodynamic equations we make the same assumption as in the master equation, that the creation and annihilation events of $\sigma^z$ operators in the operator string, due to CNOT gates, are dilute and uncorrelated. 
In a correlated collision two $\sigma^z$ operators that were generated by a CNOT gate collide again, resulting in the annihilation of one of them.
In other words, a correlated collision yields a closed loop of world lines of two initially generated $\sigma^z$ operators.
The hydrodynamic equation is valid when the probability of correlated collisions is negligible, i.e. much smaller than the probability of the uncorrelated collisions where the world lines do not form closed loops. 
Below we estimate the probability of correlated collision in the ballistic and diffusive circuits separately.

\emph{Ballistic case $(p = 1)$: }
After two $\sigma^z$ operators are created by a CNOT gate, they start propagating in opposite directions under the action of SWAP gates. 
The probability that this CNOT gate generated a correlated collision is then equal to the probability that the two operators undergo a single collision each before meeting again, which is of order unity, times the probability that the two operators collide upon meeting, which is $r$. Therefore, in the limit $r\to0$ the probability to generate a correlated collision event vanishes, and the hydrodynamic description is justified.

\emph{Diffusive case $(p < 1)$: }
In this case, in addition to CNOT gates, backscattering events can be generated by missing SWAP gates.
We consider the limit $1 \ll 1/(1-p) \ll 1/r$,
so that diffusive backscatterings dominate. We ask once again what is the probability for two $\sigma^z$ operators to meet and collide with each other before colliding independently. 
Recall that the typical time between consecutive backscatterings due to missing SWAP gates is $\tau_p\sim (1-p)^{-1}$. Therefore, the probability for the two $\sigma^z$ operators to meet again without undergoing a collision is given by $p_{\rm meet} \sim (1-r)^{2\tau_p}$. 
In the regime $\tau_p\ll\tau_r$ multiple meetings of the operators can occur before a CNOT gate acts on them. Summing over all such trajectories we obtain that the probability to generate a correlated collision is $p_{\rm meet} \left[ \sum_{n=0}^\infty (1-r)^n p_{\rm meet}^n \right] r \sim (1-p)$.
In other words, in the regime $1-p \gg r$, the validity of the hydrodynamic equation is limited by smallness of $1-p$.

\subsubsection{Hydrodynamic equations for the ballistic circuit}

In the ballistic circuit, the OTOC density $C(x,t)$ decouples into left and right moving densities, $C_L(x,t)$ and $C_R(x,t)$, respectively.
Hydrodynamic equations for the evolution of these densities are given by,
\begin{align}
    \frac{\partial C_R}{\partial t} & = &  -\vBz \frac{\partial C_R}{\partial x} + r \big[C_L (1-C_R)-C_L C_R\big] \nonumber \\
    \frac{\partial C_L}{\partial t} & = &  \vBz \frac{\partial C_L}{\partial x} + r \big[C_R (1-C_L)-C_L C_R\big].
\label{eq:kinetic_ballistic}
\end{align}
Note that operators on neighboring sites have opposite propagation directions. Hence, a right moving operator can be generated only from a left moving one, and annihilation occurs only when left and right moving operators meet - in this case one of them (the one corresponding to the control qubit) is annihilated.

We now argue that these coupled equations admit a traveling wave solution, with no front broadening.
We start by considering a solution of the form $C_{L,R}=C_{L,R}(x-\vB t)$, which turns the set of equations in~\eqref{eq:kinetic_ballistic} into coupled non-linear ordinary differential equations. There are two fixed point solutions for these equations, $C_{L,R}=0$ and $C_{L,R}=1/2$. Performing a linear stability analysis around these fixed points, we find that while the former point is a saddle point and thus unstable, the latter one is a stable solution, as long as $\vB>\vBz$. While this suggests that multiple velocities are possible for the propagation of the wave, as discussed in Refs.~\cite{Aleiner2016combustion,XuSwingle2019}, the physical velocity corresponds to the lower bound on the allowed velocities. 
Note however that the limit $\vB\to\vBz$ is singular, signaling a breakdown of the continuum limit. Hence, we expect the velocity of propagation to be $\vB=\vBz+v_\epsilon$, with the cutoff for $v_\epsilon$ being set by lattice-scale microscopic parameters.

\subsubsection{Hydrodynamic equation for the diffusive circuit}

In the diffusive case the total OTOC density $C(x,t)$ satisfies
\begin{equation}
    \frac{\partial C}{\partial t} =  D \frac{\partial^2 C}{\partial x^2} + r \big[C (1-C)-C^2\big],
\end{equation}
where $D$ is the diffusion constant.
This differential equaion is known as the FKPP equation~\cite{Fisher1937,Kolmogorov1937} which, similarly to the equations for the ballistic case, admits a traveling wave solution. 
Following a similar analysis to the one outlined in Ref.~\cite{Aleiner2016combustion}, we find that the physical velocity for the propagation of the wave in this case is given by $\vB=2\sqrt{Dr}$, in qualitative agreement with the behavior seen in Fig.~\ref{fig:vB_lL}.

\subsubsection{Numerical analysis of front broadening}

\begin{figure}
    \centering
    \begin{overpic}[width=0.49\columnwidth]{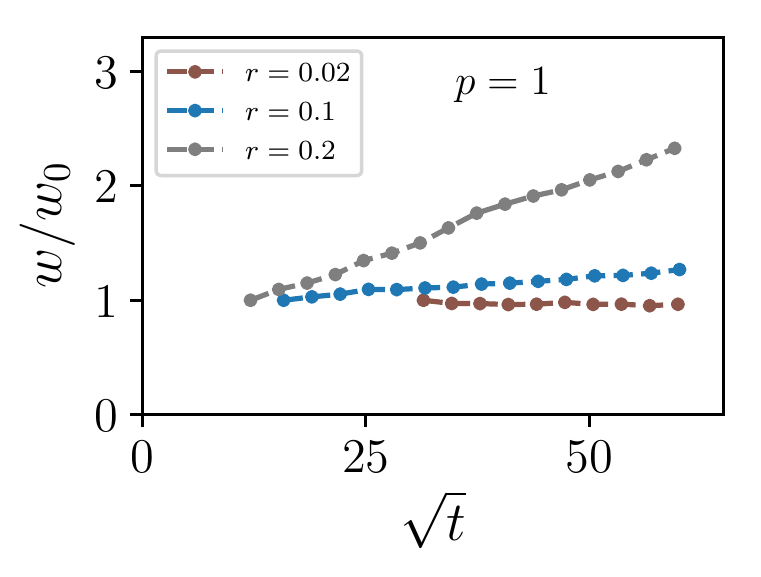} \put (0,70) {\footnotesize{(a)}} \end{overpic}
    \begin{overpic}[width=0.49\columnwidth]{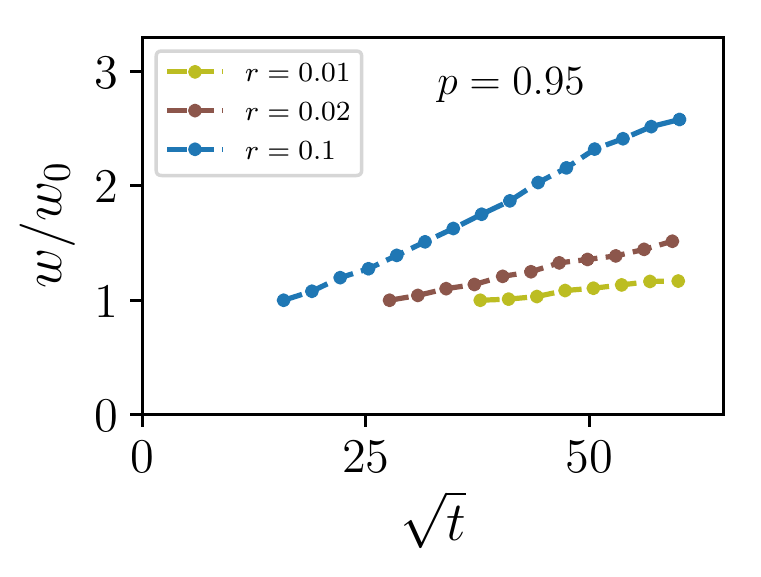} \put (0,70) {\footnotesize{(b)}} \end{overpic} \\
    \begin{overpic}[width=0.49\columnwidth]{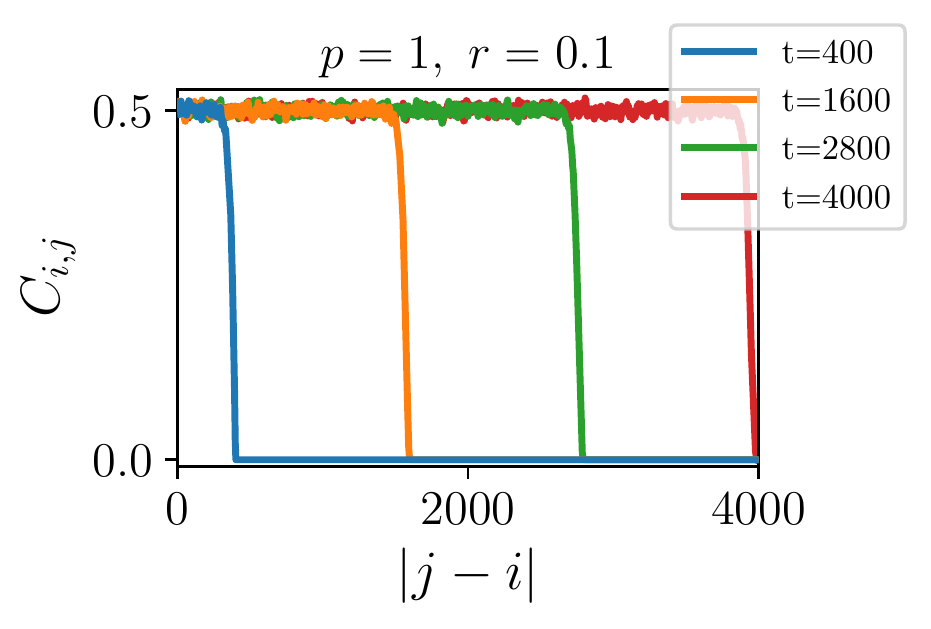} \put (0,70) {\footnotesize{(c)}} \end{overpic}
    \begin{overpic}[width=0.49\columnwidth]{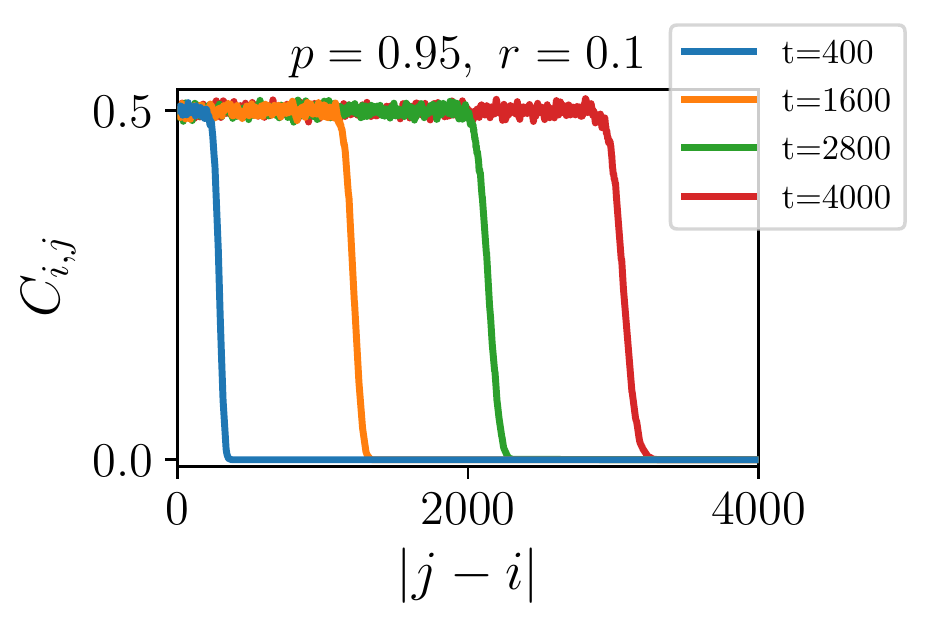} \put (0,70) {\footnotesize{(d)}} \end{overpic}
    \caption{Front broadening in the special Clifford random circuit model. (a,c) The width of the front as function of time, normalized by the width of the front at the time when the OTOC in the bulk reaches saturation, $w_0$, for $p=1$ in (a) and $p=0.95$ in (b). The local OTOC at different times for $p=1,r=0.1$ in (c) and for $p=0.95,r=0.1$ in (d).}
    \label{fig:broadening}
\end{figure}

We now present numerical results for the broadening of the front in our random circuit model.
A well-defined front develops once the OTOC reaches saturation value in the bulk, which in the main text was denoted by $C_{\rm sat}$. We define the front as the region in space where the OTOC varies between $0.2 C_{\rm sat}$ and $0.8 C_{\rm sat}$. At the time the OTOC in the bulk reaches saturation value, the front has a finite width that we denote by $w_0$ (this width is increasing with $\vB/r$). In Figs.~\ref{fig:broadening}(a,b) we plot the width $w$ normalized by $w_0$ as function of time, for $p=1$ and $p=0.95$ for different values of $r$. As can be seen, the front does not broaden (or broadens very slowly) in the limit of $p\to1, r\to0$. This is indeed the regime in which we expect the hydrodynamic equations to hold, and for a traveling wave solution without front broadening to exist. In Figs.~\ref{fig:broadening}(c,d) the front itself is shown at different times for $r=0.1$, and $p=1,0.95$.
Here, to reduce the noise, a convolution with a uniform kernel of size $10$ sites was performed (note that since the size of the kernel is small compared to the width of the front this does not alter the results).
It can be explicitly seen that while there is no broadening for the ballistic case with $p=1$, the front does broaden for $p=0.95$ for this value of $r$.

\end{document}